# A cross-platform execution engine for the quantum intermediate representation

**Elaine Wong**[1] · **Vicente Leyton-Ortega**[1] · **Daniel Claudino**[1] · **Seth R. Johnson**[1] · **Austin J. Adams**[2] · **Sharmin Afrose**[1] · **Meenambika Gowrishankar**[3] · **Anthony Cabrera**[1] · **Travis S. Humble**[4]

## Abstract

Hybrid languages like the quantum intermediate representation (QIR) are essential for programming systems that mix quantum and conventional computing models, while execution of these programs is often deferred to a system-specific implementation. Here, we develop the QIR Execution Engine (QIR-EE) for parsing, interpreting, and executing QIR across multiple hardware platforms. QIR-EE uses LLVM to execute hybrid instructions specifying quantum programs and, by design, presents extension points that support customized runtime and hardware environments. We demonstrate an implementation that uses the XACC quantum hardware-accelerator library to dispatch prototypical quantum programs on different commercial quantum platforms and numerical simulators, and we validate execution of QIR-EE on IonQ, Quantinuum, and IBM hardware. Our results highlight the efficiency of hybrid executable architectures for handling mixed instructions, managing mixed data, and integrating with quantum computing frameworks to realize cross-platform execution.

✉ Elaine Wong
wongey@ornl.gov

[1] Computing and Computational Sciences Directorate, Oak Ridge National Laboratory, Oak Ridge, TN, USA

[2] School of Computer Science, Georgia Institute of Technology, Atlanta, GA, USA

[3] Bresden Center, University of Tennessee Knoxville, Knoxville, TN, USA

[4] Quantum Science Center, Oak Ridge National Laboratory, Oak Ridge, TN, USA

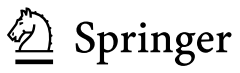

## 1 Introduction

The practice of quantum computing is developing rapidly as quantum hardware platforms and their associated programming tools begin to converge. Quantum programming integrates abstract quantum algorithms expressed in high-level quantum programming languages (QPLs) with classical logic and workflows that leverage conventional computing systems [1, 2]. The resulting hybrid programs combine quantum and classical instructions that then execute across these different hardware platforms [3–6]. Recent efforts to standardize hardware-agnostic intermediate representations (IRs) of hybrid programs expand the portability of programming across different platforms while also supporting the optimization of conventional and quantum logic [7, 8].

By contrast, variations in the capabilities and instruction set architecture (ISA) of the targeted quantum platforms require hardware-specific compilation strategies [9, 10]. These competing representations of the hybrid program introduce a need to translate the hardware-agnostic IR to a selected hardware ISA. Addressing this challenge is pivotal toward providing a means to program a quantum computer while supporting the evaluation of program performance and behavior across different technologies.

Managing program compilation for different choices of QPLs and quantum hardware requires robust methods for translating across different levels of abstraction [11]. The IR concept is familiar in conventional computing for managing the complexity of translation by tailoring the abstraction level to the intended optimizations and this ecosystem been growing in its scope and diversity as illustrated in Fig. 1. Notably, the figure highlights several of the evolving languages built on the concept of using IR to manage language complexity and to serve as an input to subsequent execution methods. For example, OpenQASM has become widely used for expressing quantum programs, including those with limited classical instruction, that are subsequently parsed by python. Similarly, several languages such as cirq, pytket, and qiskit have developed hierarchical IR's to enable library based execution methods, while there are a growing diversity of language dialects. The role of the IR becomes even more relevant as computational paradigms include quantum computing and expand to specialized dialects capable of managing quantum-specific abstractions. Additionally, a quantum IR must encapsulate the probabilistic nature of quantum operations and their outcomes alongside conventional computing operations to enable hybrid execution. These features introduce additional concerns for optimizing program structure and data usage.

To this end, Microsoft, Oak Ridge National Laboratory (ORNL), and an alliance of other institutions joined forces to specify the quantum intermediate representation (QIR) as an extension of the previously defined LLVM IR with the purpose of expressing quantum programming instructions. QIR features a standardized hardware-agnostic language by which to express and reason about hybrid algorithms, and today, a consortium of stakeholders curates the specification to provide a separation between concerns of program specification and its optimal implementation [7]. QIR also ensures integration with conventional computing operations via the adherence to the LLVM IR, which was developed for the robust translation of classical programs [12]. By design, QIR may be parsed, lowered, and executed across a variety of different hybrid hardware platforms using a set of methods that represent core capabilities of a quantum computing system including, but not limited to, state preparation, quantum gate operations, and measurement processes. The ability of QIR to leverage already well-established capabilities of classical compilation allows a natural path for studying and understanding a way forward in quantum programming and compiler development.

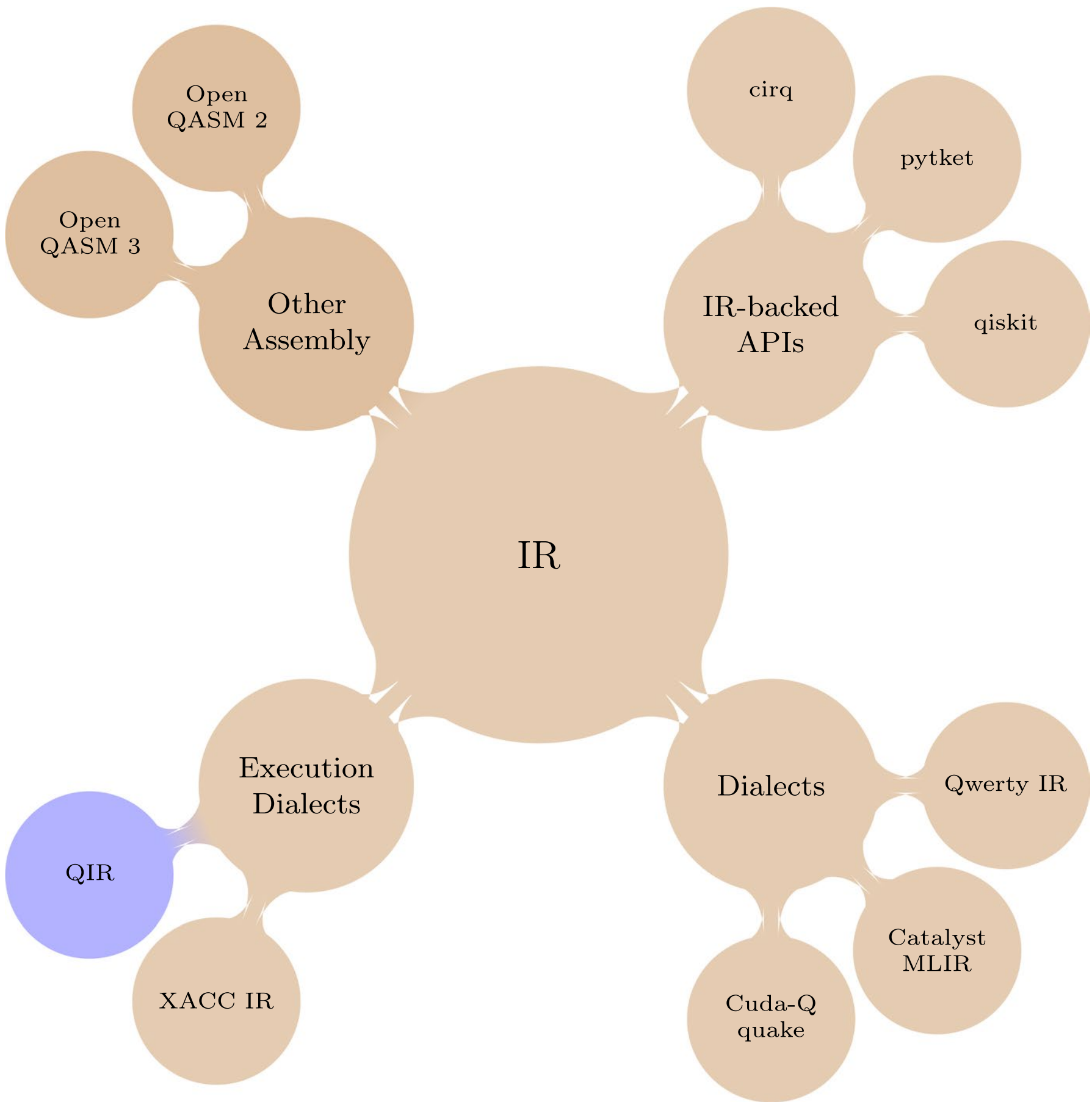

**Fig. 1** A snapshot of 'subsets' of IRs contributing to the current quantum IR ecosystem. This list is not meant to be exhaustive and many multi-level IR (MLIR) dialects may be lowered to those shown. The middle IR bubble represents the overarching node, which branches into different 'classes' of IR distinguished by their intended modus operandi: execution, being a dialect for specific purposes (such as code optimization), enabling API support, or serving as another assembly language. The outermost nodes presents as representative examples in these classes. In particular, we highlight (in blue) that QIR plays a role as the de facto standard between quantum programming languages (higher abstraction dialects) and execution dialects (assembly languages) that can potentially bridge the gap between quantum programming languages and a quantum processing unit

While the hardware-agnostic expression of QIR enables program portability, the execution environment still must issue instructions that are understood by the target platform. To this end, subsets of the QIR specification were designed to describe implemented methods which match the capabilities of a selected runtime system. Target-specific compilation then supports efficient generation of hardware-specific program instructions for the corresponding methods during platform execution. In other words, QIR specifies the way that language-specific compilers link with these platforms. Figure 2 provides a listing of object types and their purpose in QIR. This list identifies object candidates for inclusions in a *profile*, a coherent subset of functionality that a specific target can support [7]. The flexibility of the profile ensures that QIR serves a broad variety of quantum computing platforms and facilitates interoperability. The ability to identify such objects in a QIR program is intended to enable the execution environment to parse the code and distinguish between quantum and classical instructions. `Qubits` and `Results` are represented as pointers to opaque LLVM structure types. The former is so that register locations may be distinguished from other value types and the latter allows each platform implementation to provide a structured definition for measurement outcomes. The remaining objects are described further in the Appendix together with a sample program to illustrate where quantum instruction sets (QIS) and runtime functions are declared and specified.

Here, we demonstrate the effective management and execution of QIR within a hybrid software stack. We build upon the infrastructure underlying QIR to develop an execution environment that is based on the LLVM just-in-time execution engine [12]. We then use the XACC programming framework to realize hardware-specific implementations of QIR profile methods [3]. XACC provides just-in-time compilation for interpreting the quantum-specific instructions that, when coupled with LLVM, realizes a cross-platform environment for execution. Thus, our development of this environment facilitates the execution of QIR programs while maintaining the modularity of the hybrid software infrastructure. Fundamentally, our results provide a demonstration of how to execute hybrid QIR programs across a diverse range of hardware backends by integrating with the LLVM infrastructure for compilation and execution.

| **Section of QIR** | **Purpose** |
|---|---|
| type definitions for `Qubit` and `Result` | define opaque pointers |
| global constants (optional) | stores string labels |
| entry point definition | defines the quantum kernel |
| declarations of QIS functions | specify quantum interface |
| declarations of runtime functions | specify runtime interface |
| one or more attribute groups | annotate functions |
| module flags | specify high-level circuit properties |

**Fig. 2** The basic contents the QIR specification defined within the LLVM framework. An example use of each object is provided in the sample QIR file of the Appendix

We note that this paper is not a benchmarking paper, but rather a paper that highlights execution capability of an evolving specification as the community rallies around what makes the most sense in the development of a standard intermediate representation, and the design principles behind how to make that happen, together with a corresponding software repository [13].

**Summary of Main Contributions**:

A standalone library implementation that leverages the LLVM execution engine to be used for executing hybrid classical-quantum programs that follow the QIR specification.

Showcase of cross-platform execution that includes quantum hardware and simulators.

Portable, modular, and flexible implementation that is frontend agnostic, allowing for integration into QC-HPC frameworks and plug-in to multiple vendor backends.

The presentation is structured as follows: In Sect. 2, we discuss the state-of-the-art in software development for the QIR toolchain in this rapidly developing field, and the role we wish to address in this space. Following this, Sect. 3 provides an overview of the architecture and the operational logic of the tool we have developed, including its integration with the LLVM compilation flow and its core components, such as the `executor` class and app interfaces. Section 4 showcases how our tool facilitates the execution of QIR across various scenarios, from initial user input to final execution on quantum hardware or simulators. We present the potential implications of our tool for the quantum computing workflow in Sect. 5, touching on both the challenges and opportunities it presents for future research and development.

## 2 Related work

As an intermediate representation, QIR is generated by the transformation of a higher-level language or other IRs. Several compilers output QIR including the QCOR language and compiler [14, 15]. The QCOR compiler extends the Clang compiler to lower an input program authored in OpenQASM3 to QIR [14]. Clang itself is a frontend to the LLVM toolchain for C-based programming languages and the basis for compilation of the hybrid program into a QIR instance. Other toolchains for generating QIR include the python-based pyQIR by the QIR Alliance [16], CUDA-Q from Nvidia [17], QBraid's qbraid-qir software [18], and Microsoft's qiskit-qir tool [19]. Recent work has also demonstrated the transformation of other QPLs into QIR using LLVM's multi-level intermediate representation (MLIR) [20]. MLIR promotes easy and efficient extensibility and reusability of IR components, and several efforts are underway to develop non-trivial MLIR passes for quantum instructions [21, 22].

Despite these essential advances in developing a QIR toolchain, a gap in the integration of the hybrid execution environment remains. For example, the QCOR language provides a C/C++-style programming model that relies on embedded quantum kernels to isolate quantum computing operations. QCOR then uses the service-oriented model from XACC to translate these quantum-only instructions to a hardware-specific executable. Importantly, XACC has its own intermediate representation and does not execute QIR, as LLVM components are not integrated into its framework. This isolation bypasses the benefits of using QIR for hybrid program representations. Other toolchains that support quantum program execution have focused on hardware-specific ecosystems and backends and lack the portability for execution on multiple platforms [23–26].

Recently, the execution tool qir-runner from the QIR Alliance provided an initial example for how to realize hybrid execution within the LLVM ecosystem [27]. Qir-runner is a rust implementation of the QIR base profile that is executed using Inkwell [28], a wrapper with a strongly typed interface that safely exposes LLVM. We note that qir-runner is currently limited to executing QIR on a bundled numerical simulator, and we describe below a generalization of this concept for providing cross-platform execution. In particular, we show how a programming flow from the user to the hardware-specific platform is realized without requiring physics or knowledge of how the target hardware works. Furthermore, as of this writing, another QIR Alliance member–Quantinuum–built an in-house implementation of an execution engine for QIR that targets its own hardware. As far as we are aware, this is the first vendor enabling such capability.

In the following section, we explore how to effectively parse and execute QIR by linking with LLVM's execution engine. We emphasize a modular design which supports extensive customization and integration with a variety of hardware platforms. We show how we manifest this in the form of a software tool written in C++.

## 3 QIR execution engine

Building upon our previous discussion of cross-platform execution, this section dives into the QIR Execution Engine (QIR-EE). Figure 3 presents a conceptual diagram that illustrates the steps and interactions within QIR-EE, aiding in understanding how the engine integrates various functionalities essential for quantum computing. The source code and documentation for QIR-EE, accessible online [13], offers further insights into its operation.

To facilitate collaborations, QIR-EE is designed as a library that can be integrated into higher-level scientific applications as frontends, and extended via different quantum applications as backends. For example, a developer working with a new superconducting qubit processor could implement their backend by defining how each QIS function should be translated and executed on their hardware. This simplifies the development process and accelerates the adoption of new quantum technologies. Our modular design, therefore, fosters innovation and ensures that QIR-EE remains a versatile and powerful tool for quantum computation. In the following subsections, we describe in greater detail the components corresponding to each step in the path toward execution.

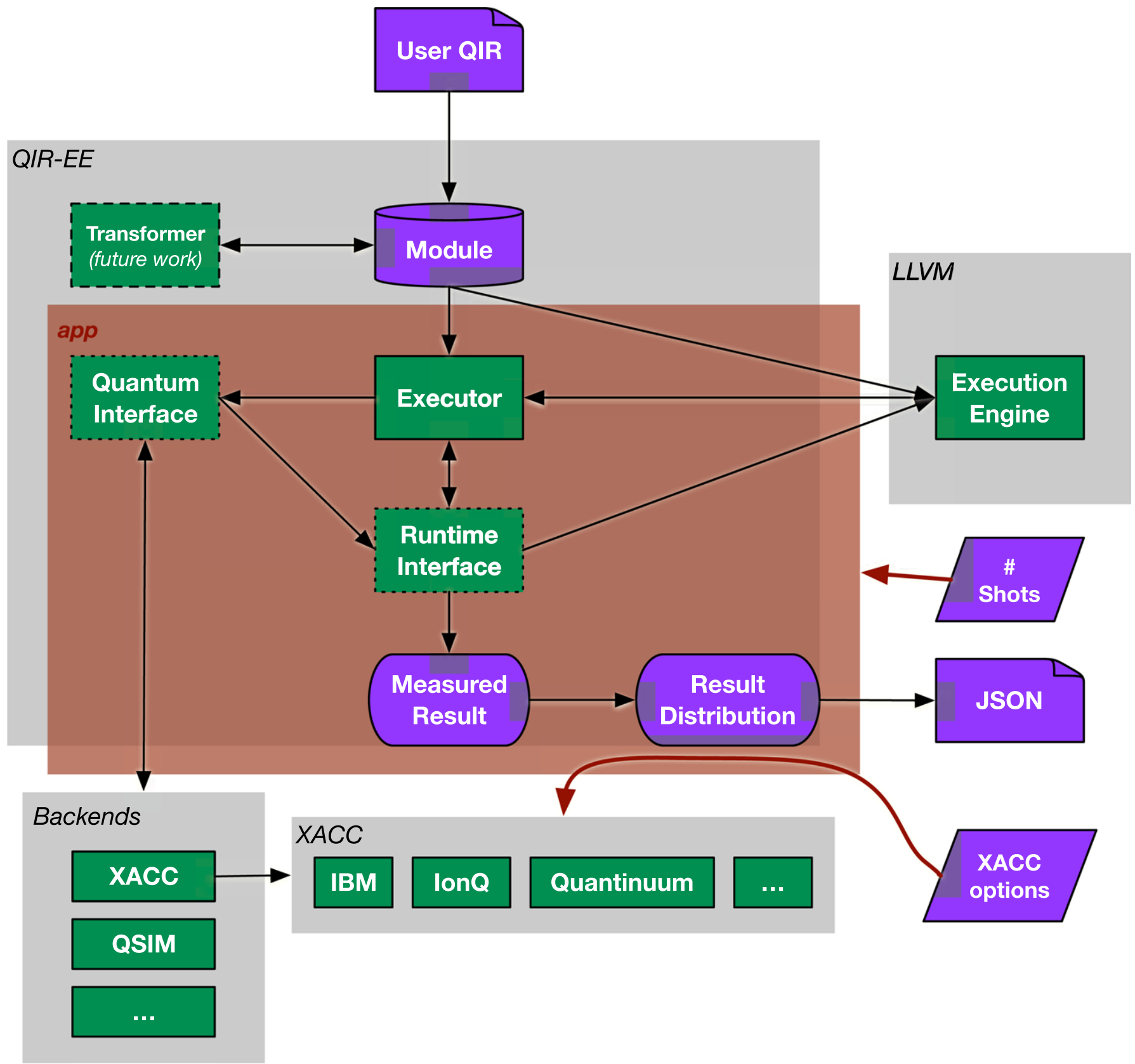

**Fig. 3** The design of QIR-EE and its relation to other components of the QIR toolchain. Taking the QIR file as user input together with the number of shots, we can view the workflow from top to bottom as follows. ❀ The module wraps a shared LLVM context that exceeds the lifetime of any IR. ❀ The executor then generates the bindings to QIR functions and manages information flow between interfaces and the LLVM execution engine. ❀ The interfaces help to separate the quantum and runtime aspects of QIR. ❀ The qir-app (example where app is the XACC framework is represented in the red box to show how it connects the different pieces) provides an additional interface for the framework(s) that enables linking to the desired hardware(s). ❀ QIR-EE also manages the measured result and compiles it into a result distribution to format into the output

### 3.1 QIR modules

QIR-EE parses and interprets QIR-formatted textual `.ll` LLVM IR programs, which can be generated using tools such as pyQIR, cudaq-quake, and qiskit-qir. Examples of such files can be found in the Appendix A.

The first step to executing QIR programs is to load the textual LLVM `.ll` file into a QIR-EE Module. This class, which wraps an LLVM module, extracts key attributes defined by the QIR standard, including the required number of qubits and classical registers necessary for execution. The module identifies the "entry

| Essential | Parameterized | Entangling | Phase | Other |
|---|---|---|---|---|
| $H$ | $R_X(\theta)$ | $CNOT, CCNOT$ | $S, S^\dagger, SY$ | $SWAP$ |
| $X$ | $R_Y(\theta)$ | $CX, CY, CZ$ | $T, T^\dagger$ | |
| $Y$ | $R_Z(\theta)$ | $ZZ, XX$ | | |
| $Z$ | $R_{ZZ}(\theta)$ | | | |

**Fig. 4** A summary of the gate binding implementations currently available in QIR-EE

```
1  qiree::QuantumRuntime* q_interface_;
2
3  void QIREE_QIS_FUNCTION(cnot, body)(std::uintptr_t arg1,\
4  std::uintptr_t arg2)
5  {
6    return q_interface_->cnot(Qubit{arg1}, Qubit{arg2});
7  }
8
9  #define QIREE_BIND_QIS_FUNCTION(FUNC, SUFFIX) \
10   bind_function("__quantum__qis__" #FUNC "__" #SUFFIX, \
11   QIREE_QIS_FUNCTION(FUNC, SUFFIX))
12 {
13   GlobalMappper bind_function(llvm_module, exec_engine);
14   QIREE_BIND_QIS_FUNCTION(cnot, body);
15 }
```

**Fig. 5** An example of 'binding' a CNOT gate, which requires two Qubit objects as input, to a QIR-adherent CNOT gate

point" that sets up the quantum circuit. It can also execute pre-processing and optimization targeting specific backends.

Because the loaded module is, in essence, a bundle of IR code blocks, LLVM-based transformations can be applied at this stage. Future work to QIR-EE will add an extension point for quantum optimizations and quantum-classical transformations, including circuit cutting [29, 30].

### 3.2 Runtime and quantum interfaces

The core principle of QIR is defining specially named functions as placeholders for quantum gates and instructions such as measurements and recording outputs.

To enable the set up of the circuits in a quantum program, functions adhering to the QIR specifications are defined through abstract interface classes, and these correspond to the gates in the "quantum instruction set" to be sent for execution. Figure 4 summarizes the gate bindings that have already been implemented by the QIR-EE developers that match these instructions. We note that it is fairly straightforward to implement other bindings, if needed. See Fig. 5 for an example of how this can be done.

The set of "runtime" functions corresponds to a second class which provides an interface between the quantum hardware and the classical readout. The underlying QIR module can interleave calls to both sets of interfaces as well as to classical computing instructions and branching. The measurement (or `mz`) function ouputs results that can then be read in and stored for further processing or recorded into a distribution to be reported in a dictionary format to be used for analysis and post-processing.

### 3.3 Executor

The `Executor` class wraps an LLVM execution engine, a just-in-time (JIT) compiler that takes LLVM IR as input. The class generates bindings that map QIR functions to the runtime and quantum interfaces. See Fig. 5 for an example of how a binding is constructed. Constructing the executor invokes LLVM's just-in-time (JIT) compilation that compiles the QIR code for the local classical processor.

This executor's "call" `operator()` is given quantum and runtime interfaces as arguments, and it launches the execution by dispatching QIR calls to those wrappers. A single executor can be reused multiple times over the course of a program. One use case is to run the entry point multiple times in a simulator where each execution is a single "shot". A second use case might be to reuse the executor with different quantum interfaces, where one interface might be a pre-processing/validation step that must succeed before the second interface launches the code on quantum hardware.

### 3.4 Backend implementations

QIR-EE, highlighted in the gray box in Fig. 3, shows how the core implementation is agnostic to any backends. To establish a link to enable execution on available simulators or hardware, an app serves as a wrapper to manage the execution and collect the results. As of this writing, QIR-EE includes two sets of implementations for quantum backends, generally represented as `qir-app` in the QIR-EE workflow as depicted in Fig. 3.

We utilize the XACC programming framework [31] to manage the backend with access to quantum computing system hardware provided by the Quantum Computing User Program at Oak Ridge National Laboratory [32]. Further enhancing this modular approach, the usage of XACC allows these custom backends to leverage advanced just-in-time (JIT) compilation techniques. This enables the dynamic adaptation of QIS functions and various quantum gates to the specific requirements of the hardware. In other words, through XACC, backends can optimize the execution of these quantum operations, ensuring optimal performance and compatibility across different quantum processors.

A second backend app wraps qsim [33], a quantum simulator developed by Google. Each invocation of the executor runs a single "shot" with this interface. This implementation can be found in [13] under `src/qirqsim`, which will build the app `qir-qsim`. The implementation of `qir-qsim` showcases how the modular component of QIR-EE works, enabling additional simulation capabilities independent of XACC.

### 3.5 Frontend execution

QIR-EE includes standalone executables to serve as a crucial interface between quantum programs and their execution on specific hardware. User-defined settings, such as the choice of hardware vendor or simulator and the number of shots, are also accommodated, with defaults currently set to IBM's aer [34] simulator and 1024 shots for `qir-xacc`. A second demonstrator executable wraps the qsim simulator. Any other hardware plug-in will work similarly.

## 4 Results and examples

To further elucidate the architectural complexities and versatility of the QIR-EE, we execute some experiments over a variety of simulators such as aer [34], Quantum++ (qpp) [35], qsim [33], IonQ's ideal simulator, the Quantinuum H1-1E emulator, as well as multiple physical backends such as IonQ's Harmony (prior to its discontinuation), Quantinuum's H1-1, and IBM's ibm-torino. These results are showcased in Fig. 7 and the data for these runs can be found in [36]. A subset of IBM's hardware yielded a successful workflow but with less than desirable results due to noisy hardware runs without error mitigation. We consider quantum phase estimation (QPE) and the quantum teleportation experiments, which pertain to the essential foundation of a high-level understanding of the QIR-EE's ability to execute complex quantum algorithms and its architectural flexibility to parse and manage quantum and classical instructions on varied experimental arrangements.

### 4.1 Quantum phase estimation

We demonstrate the execution of a prototypical hybrid QIR program using QIR-EE by validating the quantum phase estimation (QPE) algorithm. The QPE algorithm solves the following problem: *Given a unitary operator $U$ with an eigenstate $|\Psi\rangle$ with an unknown eigenvalue $e^{i2\pi\varphi}$, estimate the phase $\varphi$* [37].

The QPE algorithm exploits the direct access to a $n$-qubit unitary operator $U$ and its eigenstate $|\Psi\rangle$ to derive the $k$-bit estimate of $\varphi$. Figure 6 illustrates a quantum circuit designed to solve this problem. Initially, an ancillary register of $k$ qubits is prepared in the uniform superposition state while the primary register holds the eigenstate $|\Psi\rangle$. The quantum circuit employs a series of controlled operations, $C-U^{2^j}$, where $j$ ranges from 0 to $k-1$, and concludes with the application of the inverse quantum Fourier transform (QFT$^\dagger$). Measurement of the ancillary register subsequently provides a $k$-bit approximation to the phase $\varphi$.

For instance, in our demonstration (shown in Fig. 7), the QPE circuit from Fig. 6 designed explicitly for a six-qubit system to estimate $\varphi = 1/3$ with a precision of five bits ($n = 1, k = 5$).

Afterward, a change of basis is performed using the QFT$^\dagger$ to rearrange the extracted bits into the desired order. This critical function, which involves a sequence of groups of controlled phase gates and a Hadamard gate at each step, is programmatically achieved using `for` loops. Although not detailed in the figure, this setup is essential for the algorithm's success.

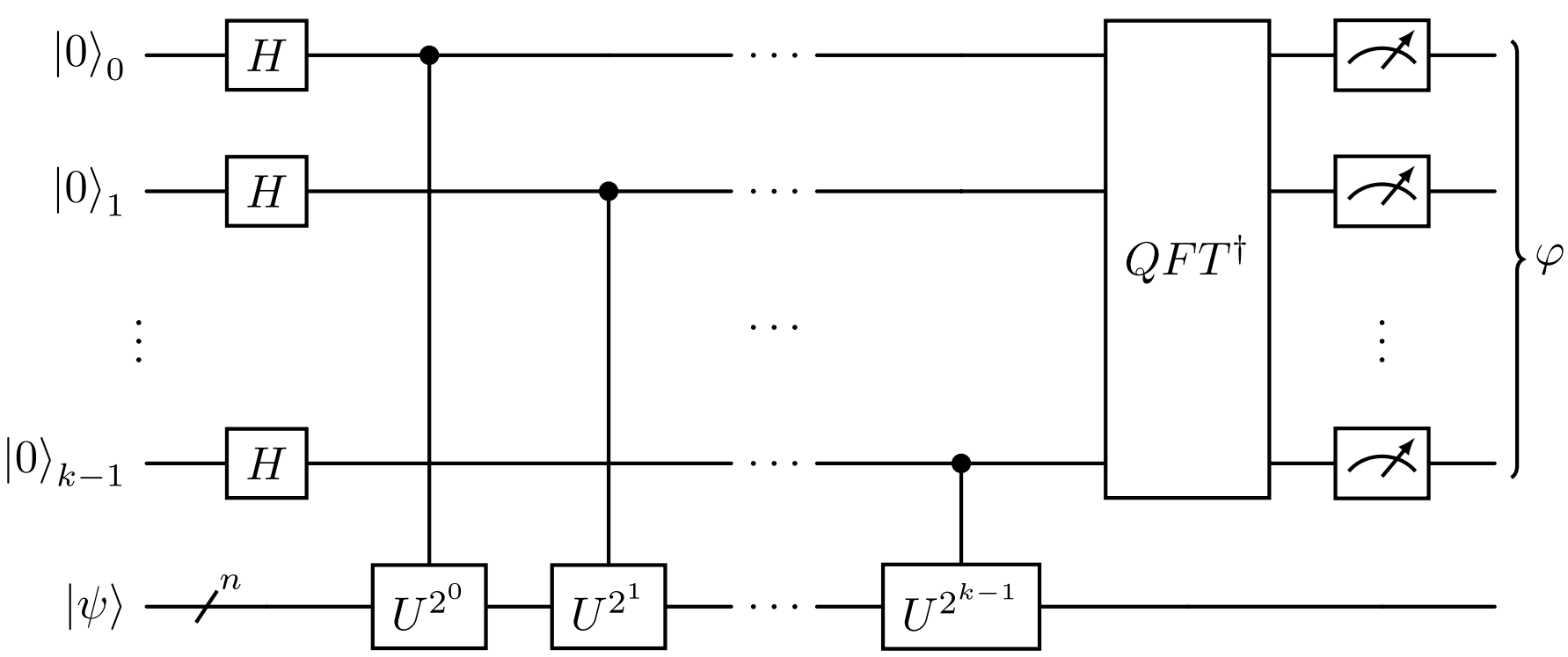

**Fig. 6** Quantum circuit representative of the quantum phase estimation (QPE) algorithm to estimate the leading $k$-bits for $\varphi$

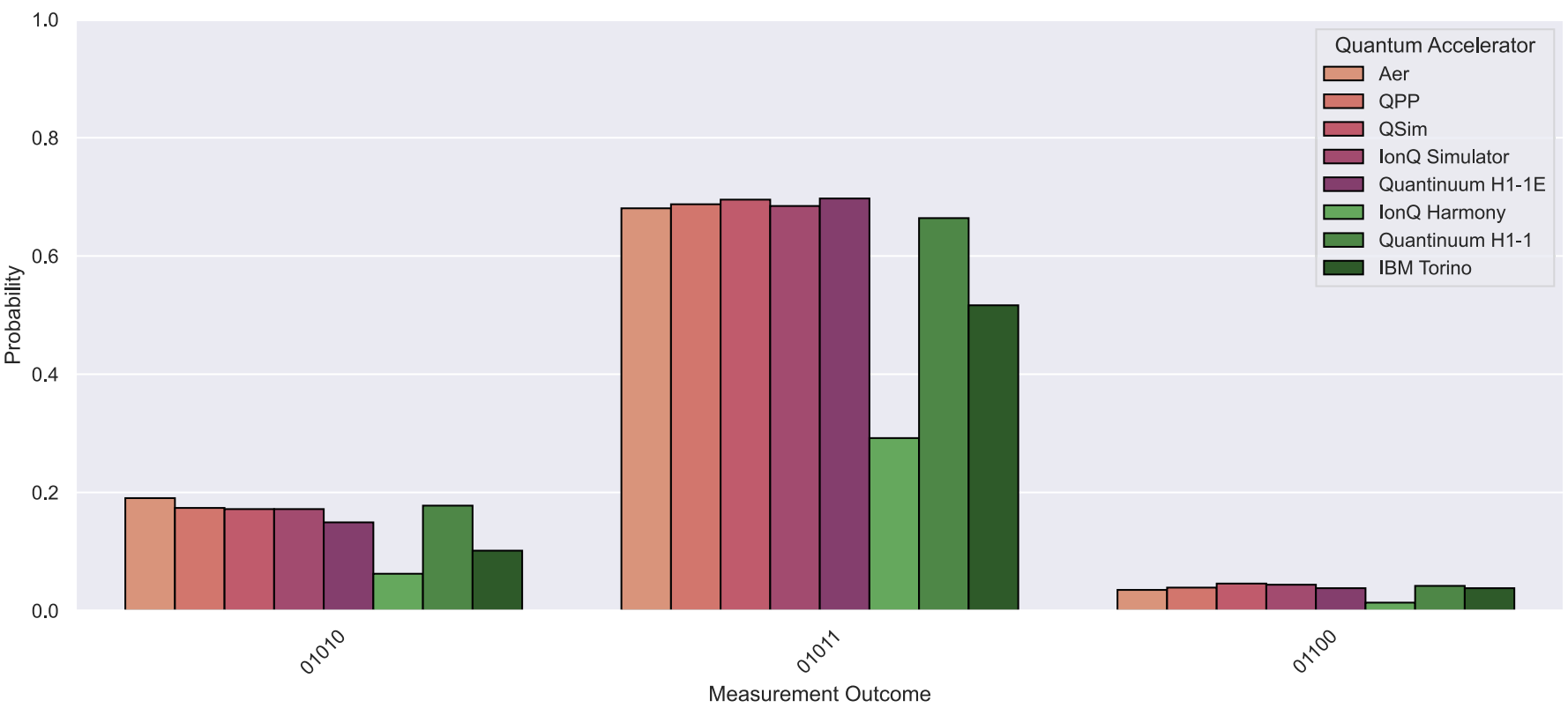

**Fig. 7** Cross-platform execution results of quantum phase estimation (QPE). This figure displays the outcomes of using QPE to estimate $\varphi = 1/3$ with a precision of five bits across various platforms. Results from simulators are depicted in shades of red, whereas those from physical quantum hardware are in shades of green. Each experiment was run with 1024 shots, and the figure highlights the *top three* measurement outcomes by probability for each experiment. We acknowledge that there are other results as well as some undesired outputs due to noise. However, the purpose of this figure is to illustrate the successful end-to-end execution of our workflow with a variety of backends/simulators. The full detailed data from these experiments are available in [36]

Subsequently, the pyQIR tool [38] parsed the Python code and generated QIR `.ll` file, which was then used for input into the QIR-EE. In this process, pyQIR unrolled the `for` loops, translating them into a final circuit configuration comprising 222 gates within the main program block and incorporating five measurement and recording commands. However, it is essential to note that even if pyQIR had preserved the `for` loops in their original form, our QIR-EE implementation would still be capable of handling them effectively. This is due to the hybrid capabilities of QIR-EE, which is built on LLVM architecture that inherently supports loop structures.

Using XACC as a means to expose the chosen accelerators, QIR-EE effectively parses and extracts quantum gate instructions from the QIR file. It then applies a global mapper to translate these instructions into an IR that XACC can process, which is an IR that is described to be polymorphic and amenable to low-level quantum program optimizations [3]. The instance of QIR-EE that is created together with XACC allows the instructions to be added as XACC's IR via the `Executor` as the QIR is parsed. This integration allows the program to be executed across various backends that are compatible with XACC, demonstrating the versatile backend support of QIR-EE. Figure 7 showcases the results from executing the QPE algorithm via the QIR program on different backends. These results highlight the efficacy of QIR-EE in conjunction with the XACC runtime, facilitating diverse quantum computational environments.

To illustrate the effects of QIR-EE on program execution, we also leverage the QPE algorithm to reveal the impact increasing the size of QIR files may have on runtime, by scaling up the qubits used in the algorithm (this is typically done to achieve better precision of the computed phase), and thus increasing the lines of code produced by the algorithm and file size. Figure 8 shows how the majority of the runtime is burdened on the backend as the file scales, while QIR-EE efficiently coordinates the preparation of the file for execution with decreasing relative cost. In

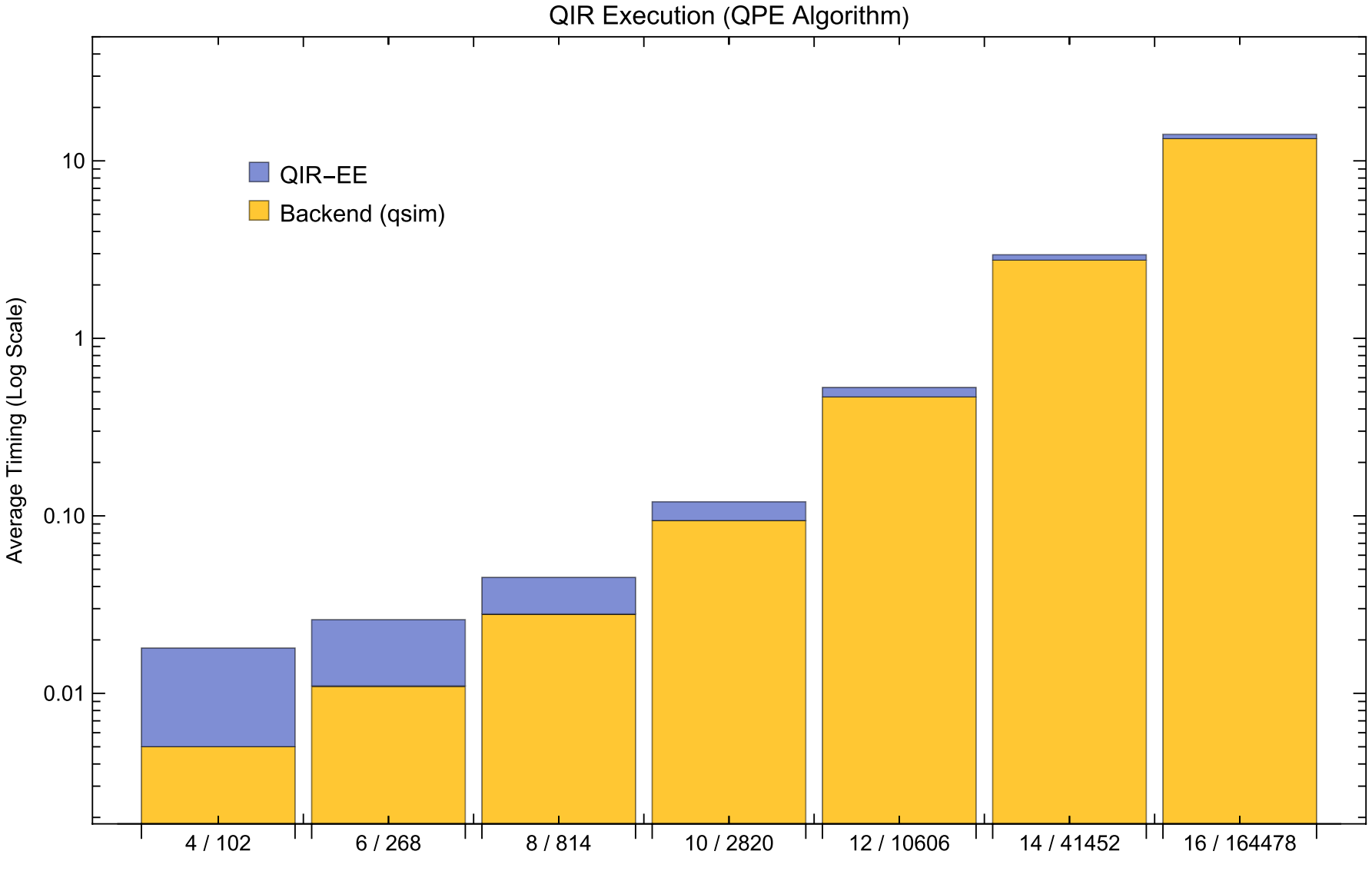

**Fig. 8** This series of timings illustrate that the effects of QIR-EE becomes negligible as a QIR program scales. For illustrative purposes, we use the QPE algorithm discussed in this section and increase the number of qubits to generate the lines of code in the QIR file to be executed on qsim backend

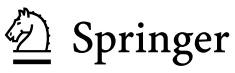

each case, the ease of using QIR-EE to run these tests is reflected in the execution of a single command line

```
qir − qsimqpe_n_qubits.ll − shotss
```

where the backend, QIR program file, and number of shots `s` are parameters that can be modified.

### 4.2 Dynamic circuit execution

In quantum computing, mid-circuit measurements are utilized for the management of quantum algorithms and to enhance error correction, contributing to the fidelity and robustness of quantum computations [39]. A mid-circuit measurement involves observing the state of selected qubits during an ongoing computation without terminating it. These measurements facilitate error syndrome detection by allowing for the assessment of ancillary qubits to determine potential errors on computational qubits without collapsing their quantum states [40, 41]. Additionally, mid-circuit measurements support fault-tolerant protocols by enabling dynamic error correction, where corrections are applied based on feedback from the system, thus allowing computations to proceed with improved accuracy [42].

An application of mid-circuit measurements and feed-forward mechanisms is demonstrated in the quantum teleportation experiment [43, 44], where the quantum state of a qubit is effectively recreated at another qubit. This process involves the destruction of the original qubit's state upon measurement, consistent with the quantum no-cloning theorem [45], and the use of classical information combined with a quantum resource to establish the state at a new qubit.

Furthermore, even if programmers do not write mid-circuit measurements explicitly, compilers may introduce mid-circuit measurements as an optimization to reduce qubit usage. For example, an optimizing compiler could split the oracle in the dynamic Bernstein-Vazirani algorithm into smaller pieces joined by mid-circuit measurements to reduce the algorithm into a form that always uses two qubits regardless of the oracle size, as shown in Fig. 9 [46].

Further illustrating the practical implementation of these principles, the capabilities of the QIR-EE to incorporate and manage feed-forward results are exemplified through its integration with `qsim` [33]. This enhancement demonstrates that QIR-EE can be adapted to interface with any quantum simulator, leveraging existing technologies to execute quantum instructions seamlessly across different platforms. While the integration capabilities of QIR-EE are robust, effectively managing the dynamic elements of quantum circuits presents ongoing challenges.

Although current hardware and simulators are generally not equipped to handle such advanced control flows, QIR-EE addresses these challenges with a customized execution backend runtime. This runtime architecture is designed to interface effectively with any supporting technologies that enhance feed-forwarding capabilities and dynamic circuit management. The flexibility and adaptability of QIR-EE are demonstrated by the successful execution of the quantum teleportation algorithm using qsim, the circuit details of which are depicted in Fig. 9 and the corresponding QIR programs outlined in the Appendix. Results from simulations, which utilize dynamic feed-forwarding during computations, are documented in [36], showing that the same experiments can be run with consistent outcomes across different simulated or physical quantum computing platforms.

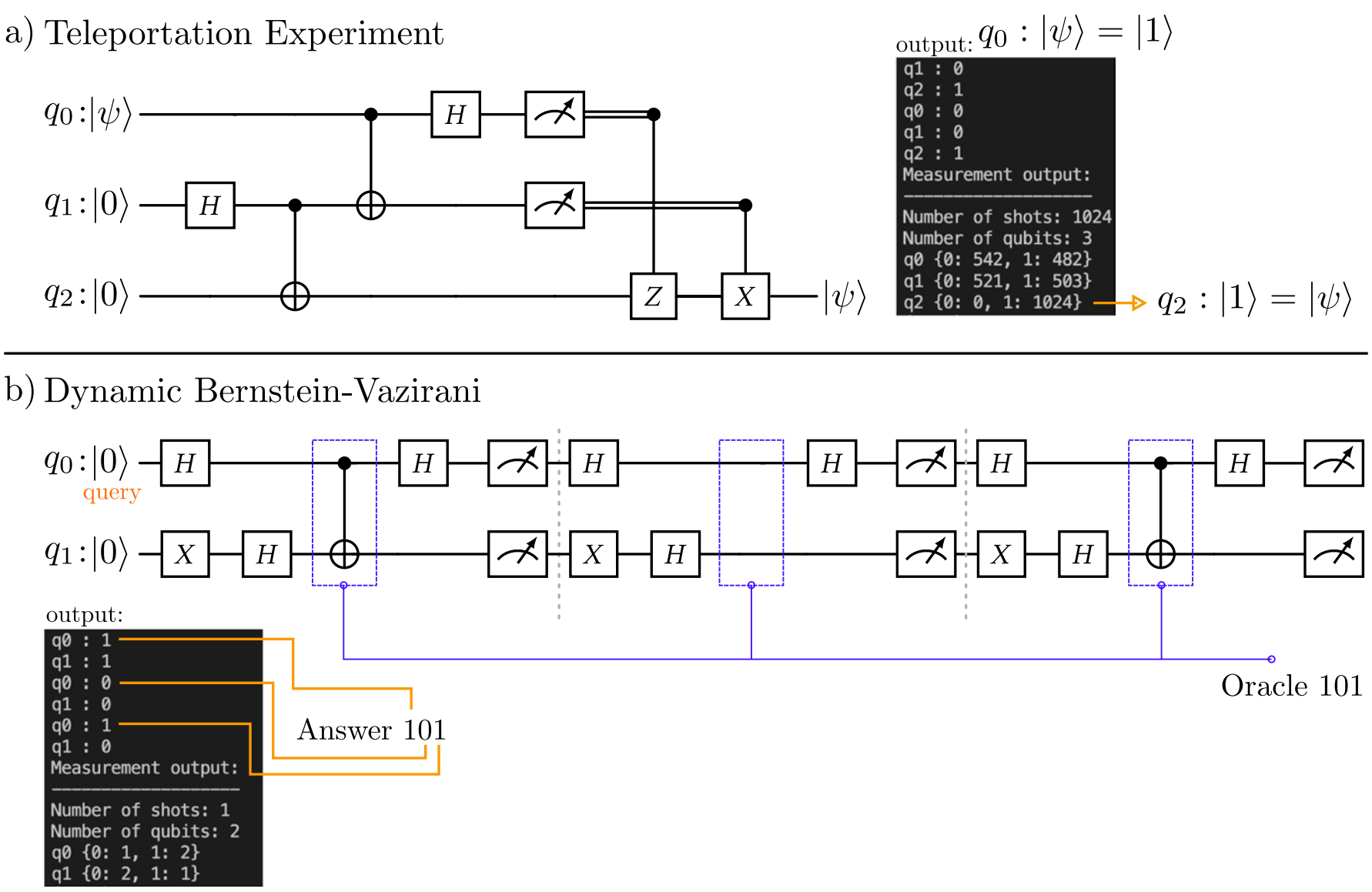

**Fig. 9** Illustration of quantum experiments executed using QIR-EE with qsim. (a) Teleportation experiment: The quantum circuit employs three qubits, where the state $|\psi\rangle = |1\rangle$ is initially on qubit $q_0$ and is teleported to qubit $q_2$. We present the quantum circuit, and the output data that shows the number of times the state $|1\rangle$ is observed in 1024 shots, confirming the successful teleportation of the quantum state. (b) Dynamic Bernstein-Vazirani experiment: The circuit uses two qubits with $q_0$ as the query and $q_1$ as the ancillary qubit. An oracle for the bitstring "101" influences the system, and the output measurement reflects the encoding of this bitstring in the state of $q_0$. Both experiments demonstrate the capabilities of QIR-EE in managing and executing complex quantum computations

## 5 Discussion and outlook

We have presented QIR-EE to parse and execute hardware-agnostic hybrid QIR programs on a variety of specific quantum computing platforms. The purpose of QIR-EE is to combine the execution of conventional and quantum instructions by using a modular design that permits extension to hybrid platforms. Within the software stack for operating a quantum computer, QIR-EE addresses the hybrid programming concerns raised within QIR and orchestrates their execution within a designated runtime, accelerator, or hardware environment. QIR-EE supports a paradigm in which each of these layers is designed independently without compromising wider portability.

Importantly, the QIR-EE framework demonstrates the synergy of executing both conventional and quantum computing elements. By leveraging LLVM IR and Execution Engine alongside the XACC programming framework, we demonstrate QIR-EE as a robust, flexible, and open-source approach to enable quantum computing system research and application development. The inclusion of the `qir-qsim` app to leverage Google's state vector simulator reveals the ability to link to other backends, moving beyond the dependency on XACC, and thus facilitating future heterogenity. We also note that certain optimizations of program code can be done at the LLVM level or at possibly a higher MLIR level, in the form of interfaces to run passes on the loaded module. Issues related to these optimizations are in the github repo [13] to be addressed.

QIR-EE addresses several challenges that arise at the intersection of the programming and execution models for hybrid systems. Foremost is the ability to parse the QIR program itself using the LLVM framework. We have demonstrated this for the latest specification of QIR and we anticipate future versions with more advanced features will be compatible as well. We have also addressed managing multiple backend targets through the extensible design of QIR-EE. We provide implementations of QIR-EE that work across a diverse collection of numerical simulators, emulators, and actual quantum computing systems.

As an interface, QIR-EE is sensitive to the varying communication patterns across vendors as well as the unpredictable latencies of remote queues. These are necessary areas of improvement for ensuring hybrid program efficiency and portablity. The ability to correctly send, wait, receive, and parse results during execution depends strongly on the hardware-specific runtime and accelerator implementation. Our demonstration of the hybrid functionality of QIR-EE through a hybrid program with conditional logic faced a similar challenge. The example highlights the current limitations of feed-forwarding programming in current quantum hardware, which is anticipated to become more sophisticated over time.

The adoption of QIR-EE for hybrid program execution depends on lowering input programs to a specified QIR profile. The latter defines the relevant hardware-agnostic QIS, and adoption requires coordination with frontend compilation to support subsequent parsing and execution. For example, the recently developed Qwerty high-level programming language and compiler provides a frontend that lowers to QIR [47, 48]. Similar QIR generators that are amenable to QIR-EE execution include pyQIR from Microsoft and Cuda-Q quake from Nvidia and other members of the alliance.

Nevertheless, the strength of QIR-EE ultimately lies in its modularity to work with any QIR-emitting software to any runtime/accelerator with access to backends, and its most straightforward application reduces to a simple workflow: *QIR* → *QIR-EE* → *Hardware*. The version of QIR-EE as reported here showcases the modular design to support a versatile, simulator-agnostic and hardware-agnostic engine, and this has been demonstrated via the two app implementations `qir-xacc` and `qir-qsim`. At the time of this writing, a third implementation `qir-lightning` in collaboration with Xanadu, gives us access to their suite of Lightning simulators [49], including lightning-kokkos, which is a high performance computing gpu focused backend. We believe that this kind of modular and open-source design for program execution will enable the development of hybrid algorithms and applications, including their profiling, analysis, and optimization, with the freedom to plug-in to any choice of backend. This flexibility could potentially allow for a smooth integration of QIR-EE with existing high performance computing frameworks and infrastructures, so that the ability to scale hybrid and heterogeneous computing paradigms can be effectively realized. We leave this for future work.

## Appendix A QIR examples

The following code snippets illustrates full QIR programs under the QIR specification v0.1 for the teleportation and the dynamic Bernstein-Vazirani example, both represented visually in Fig. 9. This code snippet can be produced from various QIR generating software. Since QIR-EE is agnostic to the user's programming language, we do not reproduce the program in the original language, but rather just show the QIR to highlight elements described in Fig. 2.

### A.1 Teleportation QIR

Qubits and measurement results are represented as pointers to an opaque LLVM structure type in lines 4-5. Lines 7-54 shows the main function module which contains branch points in lines 21, 25, 28, 37, 41, and 44 which are called by lines 19, 23, 26, 35, 39, and 42. In lines 10 and 15, Hadamard gates are applied to qubits 1 and 0, respectively. Controlled-not gates are applied with qubit 1 as the control and qubit 2 as the target in line 11, and likewise for the qubits 0 and 1 in line 13. The measurements of the qubits are performed in lines 16, 29, and 45. Lines 56-64 consists of quantum function declarations. The attributes in lines 66-68 determine the structure of the program and defines a coherent set of functionalities and capabilities that might be offered by a system via the profile setting. Lines 72-75 contains the meta data that facilitates the conveying of information about the module to LLVM's subsystems.

```
; ModuleID = 'teleport'
source_filename = "teleport"

%Qubit = type opaque
%Result = type opaque

define void @main( ) #0 {
entry:
  call void @__quantum__qis__x__body(%Qubit* null)
  call void @__quantum__qis__h__body(%Qubit* inttoptr (i64 1 to %Qubit*))
  call void @__quantum__qis__cnot__body(%Qubit* inttoptr\
  (i64 1 to %Qubit*), %Qubit* inttoptr (i64 2 to %Qubit*))
  call void @__quantum__qis__cnot__body(%Qubit* null,\
  %Qubit* inttoptr (i64 1 to %Qubit*))
  call void @__quantum__qis__h__body(%Qubit* null)
  call void @__quantum__qis__mz__body(%Qubit* null, %Result* null)
  call void @__quantum__qis__reset__body(%Qubit* null)
  %0 = call i1 @__quantum__qis__read_result__body(%Result* null)
  br i1 %0, label %then, label %else

then:                                   ; preds = %entry
  call void @__quantum__qis__z__body(%Qubit* inttoptr (i64 2 to %Qubit*))
  br label %continue

else:                                   ; preds = %entry
  br label %continue

continue:                               ; preds = %else, %then
  call void @__quantum__qis__mz__body(%Qubit* inttoptr (i64 1 to %Qubit*),\
  %Result* inttoptr (i64 1 to %Result*))
  call void @__quantum__qis__reset__body(%Qubit* inttoptr\
  (i64 1 to %Qubit*))
  %1 = call i1 @__quantum__qis__read_result__body(%Result* inttoptr\
  (i64 1 to %Result*))
  br i1 %1, label %then1, label %else2

then1:                                  ; preds = %continue
  call void @__quantum__qis__x__body(%Qubit* inttoptr (i64 2 to %Qubit*))
  br label %continue3

else2:                                  ; preds = %continue
  br label %continue3

continue3:                              ; preds = %else2, %then1
  call void @__quantum__qis__mz__body(%Qubit* inttoptr (i64 2 to %Qubit*),\
  %Result* inttoptr (i64 2 to %Result*))
  call void @__quantum__rt__array_record_output(i64 3, i8* null)
  call void @__quantum__rt__result_record_output(%Result* null, i8* null)
  call void @__quantum__rt__result_record_output(%Result* inttoptr\
  (i64 1 to %Result*), i8* null)
  call void @__quantum__rt__result_record_output(%Result* inttoptr\
  (i64 2 to %Result*), i8* null)
  ret void
}

declare void @__quantum__qis__h__body(%Qubit*)
declare void @__quantum__qis__cnot__body(%Qubit*, %Qubit*)
declare void @__quantum__qis__mz__body(%Qubit*, %Result* writeonly) #1
declare void @__quantum__qis__reset__body(%Qubit*)
declare i1 @__quantum__qis__read_result__body(%Result*)
declare void @__quantum__qis__z__body(%Qubit*)
declare void @__quantum__qis__x__body(%Qubit*)
declare void @__quantum__rt__array_record_output(i64, i8*)
declare void @__quantum__rt__result_record_output(%Result*, i8*)

attributes #0 = { "entry_point" "num_required_qubits"="3"\
"num_required_results"="3" "qir_profiles"="custom" }
attributes #1 = { "irreversible" }

!llvm.module.flags = !{!0, !1, !2, !3}

!0 = !{i32 1, !"qir_major_version", i32 1}
!1 = !{i32 7, !"qir_minor_version", i32 0}
!2 = !{i32 1, !"dynamic_qubit_management", i1 false}
!3 = !{i32 1, !"dynamic_result_management", i1 false}
```

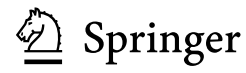

## A.2 Dynamic Berstein-Vazirani QIR

This QIR program is the code generated for the circuit in Fig. 9

```
 ModuleID = 'dynamicbv'
source_filename = "dynamicbv"

; ModuleID = 'BernsteinVazirani'
source_filename = "bv_algorithm"

%Qubit = type opaque
%Result = type opaque

define void @main() #0 {
entry:
  ; apply Hadamard to query qubit
  call void @__quantum__qis__h__body(%Qubit* null)

  ; set ancillary qubit
  call void @__quantum__qis__x__body(%Qubit* inttoptr (i64 1 to %Qubit*))
  call void @__quantum__qis__h__body(%Qubit* inttoptr (i64 1 to %Qubit*))

  ; apply CNOT for bit '1' with kickback phase on 'null'
  call void @__quantum__qis__cnot__body(%Qubit* null, %Qubit* inttoptr\
  (i64 1 to %Qubit*))

  ; correcting eigenvalue
  call void @__quantum__qis__h__body(%Qubit* null)

  ; mid-circuit measurement on first bit
  call void @__quantum__qis__mz__body(%Qubit* null, %Result* null) ;
  call i1 @__quantum__qis__read_result__body(%Result* null)

  ; reset ancillary qubit
  call void @__quantum__qis__mz__body(%Qubit* inttoptr (i64 1 to %Qubit*),\
  %Result* inttoptr (i64 1 to %Result*))
  call i1 @__quantum__qis__read_result__body(%Result* inttoptr\
  (i64 1 to %Result*))

  ; output the results
  call void @__quantum__rt__array_record_output(i64 2, i8* null)
  call void @__quantum__rt__result_record_output(%Result* null, i8* null)
  call void @__quantum__rt__result_record_output(%Result* inttoptr\
  (i64 1 to %Result*), i8* null)

  ; initialize qubits
  call void @__quantum__qis__x__body(%Qubit* inttoptr (i64 1 to %Qubit*))

  ; set ancillary qubit
  call void @__quantum__qis__h__body(%Qubit* inttoptr (i64 1 to %Qubit*))

  ; apply Hadamard to query qubit
  call void @__quantum__qis__h__body(%Qubit* null)

  ; apply Identity for bit '0'
  ; Nothing

  ; mid-circuit measurement
  ; from this we get the first bit
  call void @__quantum__qis__mz__body(%Qubit* null, %Result* null)
  call i1 @__quantum__qis__read_result__body(%Result* null)

  ; reset ancillary qubit
  call void @__quantum__qis__mz__body(%Qubit* inttoptr (i64 1 to %Qubit*),\
  %Result* inttoptr (i64 1 to %Result*))
  call i1 @__quantum__qis__read_result__body(%Result* inttoptr\
  (i64 1 to %Result*))

  ; output the results
  call void @__quantum__rt__array_record_output(i64 2, i8* null)
  call void @__quantum__rt__result_record_output(%Result* null, i8* null)
  call void @__quantum__rt__result_record_output(%Result* inttoptr\
  (i64 1 to %Result*), i8* null)

  ; initialize qubits
  call void @__quantum__qis__x__body(%Qubit* inttoptr (i64 1 to %Qubit*))

  ; set ancillary qubit
  call void @__quantum__qis__h__body(%Qubit* inttoptr (i64 1 to %Qubit*))

  ; apply Hadamard to query qubit
  call void @__quantum__qis__h__body(%Qubit* null)

  ; set ancillary qubit
  call void @__quantum__qis__x__body(%Qubit* inttoptr (i64 1 to %Qubit*))
  call void @__quantum__qis__h__body(%Qubit* inttoptr (i64 1 to %Qubit*))

  ; apply CNOT for bit '1' with kickback phase on 'null'
  call void @__quantum__qis__cnot__body(%Qubit* null, %Qubit* inttoptr\
  (i64 1 to %Qubit*))

  ; correcting eigenvalue
  call void @__quantum__qis__h__body(%Qubit* null)

  ; mid-circuit measurement on first bit
  call void @__quantum__qis__mz__body(%Qubit* null, %Result* null) ;
  call i1 @__quantum__qis__read_result__body(%Result* null)

  ; reset ancillary qubit
  call void @__quantum__qis__mz__body(%Qubit* inttoptr (i64 1 to %Qubit*),\
  %Result* inttoptr (i64 1 to %Result*))
  call i1 @__quantum__qis__read_result__body(%Result* inttoptr\
  (i64 1 to %Result*))

  ; output the results
  call void @__quantum__rt__array_record_output(i64 2, i8* null)
  call void @__quantum__rt__result_record_output(%Result* null, i8* null)
  call void @__quantum__rt__result_record_output(%Result* inttoptr\
  (i64 1 to %Result*), i8* null)

  ret void
}
```

**Acknowledgements** We acknowledge the use of resources of the Quantum Computing User Program, which is managed by the Oak Ridge Leadership Computing Facility, which is a DOE Office of Science User Facility supported under Contract DE-AC05-00OR22725.

**Author contributions** Author contributions: EW led and guided development for the work that is represented in this paper, including concept design, code development, paper writing, as well as presented content and represented the team at multiple venues (local events and international conferences) to open up opportunities for future collaborations. VLO developed the base code, contributed to writing sections of the manuscript, offered insights that shaped the conceptual framework, and collaborated closely with the team to support research objectives and to ensure the integrity of the findings. DC wrote part of the software, provided guidance on the project, and wrote part of the manuscript. SRJ developed the primary code used by the paper, contributed to writing, discussed paper layout and goals, and collaborated with team. AJA contributed to reviewing the manuscript and refactored the code used by the paper to illustrate a path to full integration with other quantum software, such as the runtime for the Qwerty programming language. SA generated initial test examples for testing and debugging the base code and contributed to the review and editing of the manuscript as well as drew the histogram in the manuscript. MG contributed to workflow development, software testing, manuscript review and editing. AC advised and provided feedback regarding the infrastructure and tooling around LLVM IR, MLIR, and QIR. TSH led the project design, analysis, and editing of the manuscript.

**Funding** Support for this work came from the U.S. Department of Energy, Office of Science, Advanced Scientific Computing Research Accelerated Research in Quantum Computing Program under field work proposal ERKJ332.

**Data availability** The datasets and outputs generated for this study can be found in the QIR-EE data GitHub repository: https://github.com/wongey/qiree-data.

**Code availability** Code corresponding to this manuscript can be found at the QIR-EE repository [13], under the ORNL-QCI organization.

## Declarations

**Ethics approval and consent to participate** All of the material is owned by the authors and/or no permissions are required.

**Conflict of interest** The authors declare no Conflict of interest.

**Consent for publication** The corresponding author has read and understood the publishing policy, and submit this manuscript in accordance with this policy.